\documentclass[preprint,showpacs,preprintnumbers,amsmath,amssymb,superscriptaddr
ess]{revtex4}

\usepackage[dvips]{graphics}
\usepackage{epsfig}

\begin{document}

\title{
Evolutionary origin of power-laws in Biochemical Reaction Network; 
embedding abundance distribution into topology
}

\author{Chikara Furusawa}
\affiliation{
Department of Bioinformatics Engineering, Graduate School of 
Information Science and Technology, Osaka University, 2-1 Yamadaoka, Suita, Osaka 565-0871, Japan
}

\author{Kunihiko Kaneko}
\affiliation{Department of Pure and Applied Sciences
Univ. of Tokyo, Komaba, Meguro-ku, Tokyo 153-8902, Japan}
\affiliation{
ERATO Complex Systems Biology Project, JST, 3-8-1 Komaba,Meguro-ku, Tokyo 153-8902, Japan}

\begin{abstract}
The evolutionary origin of universal statistics in biochemical reaction network
is studied, 
to explain the power-law distribution of reaction links 
and the power-law distributions of chemical abundances.
Using cell models with catalytic reaction network,
we find evidence that the power-law distribution in
abundances of chemicals emerges by the selection of cells with higher growth speeds.
Through the further evolution, this inhomogeneity 
in chemical abundances is shown to be embedded in the distribution of links, 
leading to the power-law distribution.
These findings provide novel insights into the nature of network evolution in
living cells.

\end{abstract}

\pacs{87.17.Aa, 87.23.Kg, 89.75.Fb}

\maketitle

Recent advances in biology have provided detailed knowledge
about individual molecular processes and their functions.  Despite its
enormous success, it is increasingly clear that the nature of
intra-cellular dynamics maintaining the living state is difficult to
be understood only by building up such detailed knowledge of
molecules, since a complex network of reactions among these molecules,
such as proteins, DNA, RNA and so forth, are essential for it.
One possible strategy to extract the nature of intra-cellular
dynamics is to search for universal laws with regard to the networks
of intra-cellular reactions common to all living systems, and then to
unravel the dynamics of evolution leading to such universal features.

Indeed, recent large-scale studies revealed two universal features in
cellular dynamics.  First, the power-law distribution of links in
reaction networks was discovered in metabolic and other biochemical
reaction networks, as is termed as a scale-free network, where the
connectivity distribution $P(k)$ obeys the law $k^{-\gamma}$ with
$\gamma \approx (2 \sim 3)$\cite{barabasi1,barabasi2,
protein_network, deletion, kepes}.  Second, the abundances of
chemicals in intra-cellular reaction were found to also exhibit the
power-law distributions, as confirmed at the levels of gene expression
\cite{Zipf, Ueda, Kuznetsov} and metabolic flux
\cite{barabasi_metabo}.  Here, the chemical abundances plotted in the order
of their magnitude are inversely proportional to their rank.

Despite the potential
importance of these universal statistical laws, however, how they are formed through evolution
and how the two laws are mutually related are still unknown.  
Here we answer these questions through analysis 
and simulations of evolution of a simple cell model, to
demonstrate that a power-law distribution in abundances of
chemicals emerges as a result of competition for
higher growth of a cell, and then this abundance inhomogeneity is 
embedded into the distribution of links, leading to the scale-free
network with hierarchical organization of reaction dynamics.
The findings provide novel insights into the nature of network evolutions
in living cells.

To determine the emergence and interrelationships of the power-laws in
chemical abundances and network connectivity through the process of
evolution, we adopted a simple model of intracellular reaction
dynamics that captures the catalytic reaction processes essential for
cell growth and division, following Ref.\cite{Zipf, KKTY1, CFKK1}.
Although this model was chosen simply to satisfy the minimal
requirements of intracellular reaction dynamics of a growing cell, it
was found to capture universal statistical behaviors as confirmed
experimentally \cite{Zipf}.  By studying a class of simple models with
these features and the evolution of the network of the reaction, we
study how the power-laws in abundances and network connectivity emerge
inevitably.

Consider a cell consisting of a variety of chemicals.
The internal state of the  cell can be represented
by a set of concentrations $(x_1,x_2,\cdots ,x_K)$, where
$x_i$ is the intra-cellular concentration of the chemical species $i$ with $i$
ranging from $i=1$ to $K$. 
Depending on whether there is an enzymatic
reaction from $i$ to $j$ catalysed by some other chemical $\ell$, the reaction
path is connected as $(i + \ell \rightarrow j + \ell)$. 
The rate of increase of $x_j$ 
through this reaction is given by $x_i x_{\ell}$, where for simplicity all
of the reaction coefficients were chosen to be 1.

Next, some nutrients were supplied from the environment by
transportation through the cell membrane with the aid of some other
chemicals, i.e., ``transporters". Here, we assumed that the transport
activity of a chemical is proportional to its concentration, and the
rate of increase of $x_i$ by the transportation is given by $D x_{m_i}(X_i-x_i)$, 
where $m_i$ is the
transporter for the nutrient $i$, $D$ is a transport constant 
and the constant $X_i$ is the concentration of the $i$-th chemical in the environment.
In addition, we took into account the changes in cell volume, which
varies as a result of transportation of chemicals into the cell from the
environment. For simplicity, we assumed that the volume is proportional
to the sum of chemicals in the cell. The concentrations of chemicals are
diluted due to increases in volume of the cell,
which imposes the restriction $\sum_{i} x_i=const$. When the
volume of a cell is doubled due to nutrient intake, the cell is assumed
to divide into two identical daughter cells.

To summarize these processes, the dynamics of chemical concentrations 
in each cell are represented as 
\begin{eqnarray}
dx_i/dt  = R_i-x_i \sum_j R_j
\end{eqnarray}
with
\begin{eqnarray}
R_i &=& \sum_{j,\ell}Con(j ,i ,\ell) \;x_j \;x_{\ell}  - \sum_{j',{\ell}'}Con(i ,j' ,{\ell}') \;x_i \; x_{{\
\ell}'} \nonumber \\
& &(+ D x_{m_i}(X_i-x_i)),
\end{eqnarray}
where $Con(i,j,\ell)$ is 1 
if there is a reaction $i+\ell \rightarrow j + \ell$, and 0 otherwise, while
the last term in $R_i$ is added only for the nutrients, and
represents its transportation into a cell from the environment.
The last term in $dx_i/dt$ with the sum of $R_j$ represents 
dilution effect due to changes in cell volume.

Of course, how these reactions progress depends on the intra-cellular reaction network.  
Here, we study the evolution of the network, by 
generating slightly modified networks and selecting those
that grow faster.  
First, $n$ mother cells are generated, where the 
connection paths of catalytic network 
were chosen randomly such that the number of incoming, outgoing, and 
catalyzing paths of each chemical is set to the initial path number $k_{init}$.
From each of $n$ mother cells, $m$ mutant cells were generated by
random addition of one reaction path to the reaction network of the
parent cell. Then, reaction dynamics were simulated for each of the $n \times m$ 
cells to determine the growth speed of each cell, i.e., the
inverse of the time required for division. Within the cell
population, $n$ cells with faster growth speeds were selected as the mother
cells of the next generation, from which $m$ mutant cells were again
generated in the same manner.

A number of network evolution simulations were performed using several
different initial networks, different parameters and various
settings. We found that all of the simulations  indicated common
statistical properties with regard to both reaction dynamics and
topology of networks. Here, we present an example of simulation results
to show the common properties of our simulations.

The rank-ordered concentration distributions of chemical species in several 
generations are plotted in Fig.1, in which the
ordinate indicates the concentration of chemical species $x_i$ and the
abscissa shows the rank determined by $x_i$. The slope of the rank-ordered
concentration distribution increased with generation, and within a few
generations converged to a power-law distribution with an exponent -1,
which was maintained over further generations. 
Or equivalently, the distribution $p(x)$ of the 
species with abundance $x$ is proportional to $x^{-2}$ \cite{henkan}.

The emergence of such a power-law by selection of cells with faster
growth speeds is a natural consequence of our previous observations. To
increase the cellular growth speed, changes in the network that enhance
nutrient uptake from the environment are favored. This nutrient uptake
is achieved by increasing the concentrations of transporters. However,
the cell will no longer be able to grow continuously if there is
excessive nutrient uptake as there will be insufficient room to
synthesize the catalysts required to ``metabolize" the nutrients and
transporters. Thus, there is a critical value of nutrient uptake at
which the cell reaches the maximal growth speed. At this point, the
power-law distribution of chemical abundance appears in the
intracellular dynamics, where the hierarchy of the successive catalytic
reaction processes is organized \cite{Zipf}. With the evolutionary process shown
in Fig.1, the nutrient uptake increases to accelerate the cellular
growth speed until further mutations in the network result in exceeding
the above critical value of nutrient uptake. Here, successive increases
in growth speed by ``mutation" to the reaction network are possible
only when the degree of enhancement of nutrient uptake by the mutation
is synchronized with increases in other catalytic
activities. Consequently, selection favors networks in which nutrient
uptake is maintained close to this critical point, where successive
catalytic reaction processes maximize the use of nutrients and form a
power-law distribution of chemical abundances.

Next, we investigated the topological properties of the reaction
networks. Although both initial network generation and reaction path
addition were random, after this evolutionary process the topological
properties of networks deviated significantly from those expected from
random networks.  The connectivity distributions $P(k)$ of chemical
species obtained from the network of the 1000th generation are plotted
in Fig.2a, where $k_{in}$, $k_{out}$ and $k_{cat}$ indicate the numbers of
incoming, outgoing and catalyzing paths of chemicals,
respectively. These distributions were fitted by power-laws with an
exponent close to -3. Thus, a scale-free network was approached through
evolution, while this power-law behavior was maintained for further
evolutionary processes. 

As shown in Fig.3, in this simple model, the evolved reaction network
formed a cascade structure in which each chemical species was mainly
synthesized from more abundant species. That is, almost no chemical
species disrupted the flow of chemical reaction from the nutrients, as
the network approached that with optimal cell growth.  It should also be
noted that the reaction dynamics for each chemical were also
inhomogeneous in that synthesis of each chemical species had a dominant
reaction path. Such uneven use of local reaction paths was also reported
previously in real metabolic networks \cite{barabasi_metabo}.

The emergence of the scale-free-type connectivity distribution by this
evolutionary process can be explained by selection of preferential
attachment of paths to more abundant chemical species. Note that the
power-law distribution of chemical abundance has already been
established through evolution. Here, the connection of a reaction path
to a more abundant chemical is more effective resulting in a change in
growth speed of a cell. For example, assume
that the change in growth speed by addition of an outgoing path from a
chemical increases linearly with its abundance $x$. This assumption is
natural as the degree of influence on the cellular state is generally
proportional to the flux of the reaction path added to the network,
i.e., the product of substrate and catalyst abundances. In this simple
case, the probability $q_{out}(x)$ of having such an outgoing path after
selection will increase linearly with $x$ even though 
the change in the network is random. 
Here, the connectivity distribution $P(k_{out})$ is obtained by the transformation 
of variable as follows.
Suppose that the probability of
selection of a path attached to a chemical with abundance $x$ is given by
$q(x)$, then the path number $k \propto q(x)$. By the transformation
$k=q(x)$, the distribution
\begin{equation}
P(k) =dx/dk p(x) =q'(q^{-1}(k)) p(q^{-1}(k))
\end{equation}
is obtained.  By applying the abundance power-law $p(x)\propto x^{-2}$, 
we obtain $P(k)=k^{-(\alpha +1)/\alpha}$ when $q(x)=x^{\alpha}$.
Consequently, a scale-free network with exponent -2 should be evolved if $q_{out}(x) \propto x$.

Numerically, we found that the probabilities $q_{out}(x)$ and $q_{cat}(x)$ were
fitted by $q(x)\propto x^{\alpha}$ with $\alpha \approx 1/2$, as shown in Fig.2b.
Then, using the above transformation
the connectivity distribution was obtained as $P(k)=k^{-3}$.
Here, it is interesting to note that the connectivity distribution
observed from real metabolic and other biochemical networks
follow the power-law $P(k) \propto k^{-\gamma} $ with $\gamma$ between 2 and 3,
as often seen in experimental data \cite{barabasi1,barabasi2}.

The probability $q(x)$ is determined through the evolutionary process.
To clarify the reason for $q(x) \sim x^{\alpha}$ with $\alpha < 1$ 
in outgoing and catalyzing paths, 
we investigated the relationship between substrate abundance $x$ and 
catalyst abundance $y$ of a path to be selected.
For this, we simulated changes in growth speeds by 
random addition of a reaction path to the network of 200th generation.
By examining only the mutants of  0.05\% of the highest growth speeds 
we found that a path with small flux is not selected 
since adding such path cannot change the cellular state enough, 
while a path with large flux is not selected also, 
since such large change destroys hierarchical structure of catalytic reactions, 
which results the decrease of nutrient intakes or exceeding 
the critical point so that the ``cell" can no longer grow.
Then by plotting the substrate abundance $x$  and catalyst abundance $y$ for the selected paths
on the $x$-$y$ plane,
we found that the fluxes of the selected paths
satisfy $\Delta < xy < \Delta +\delta$, with $\Delta= 3.8 \times 10^{-8}$ and $\delta = 4.0
\times 10^{-6}$ (data not shown).
We also found that the density of paths to be selected is almost constant 
in the above region.
Consequently, for each chemical $x$, the probability
that such a path exists is given by the probability that there is such a
partner chemical with abundance $y$, which satisfies $\Delta /x <y < (\Delta +\delta)/x$. That is,
\begin{equation}
q(x) =\int_{\Delta/x}^{(\Delta+\delta)/x} p(z) dz \approx p(\Delta/x) (\delta/x)
\end{equation}
By using the equation (1), we obtain
\begin{equation}
P(k)=\frac{-p(\Delta/y) }{(p(y)+ydp(y)/dy))y^2},
\end{equation}
with
\begin{math}
yp(y)=k.
\end{math}
Indeed, if $p(x)=x^{-2}$, the above expressions lead to $q(x) \propto
x$, as well as $P(k)=k^{-2}$.  This expression holds when the evolved
network is just at the critical point.  The evolved network is near this
critical point but there is a slight deviation, as can be seen in the deviation
from the power-law in Fig.1, for small abundance of chemicals.
Note that the asymptotic
behavior for large $k$ is given for small $y$. Then, the asymptotic
behavior for large $k$ is given by $P(k) \approx 1/((p(y)+ydp(y)/dy))$
depends on $p(y)$ for small $y$.  If the asymptotic behavior of $p(y)$
for small $y$ is given by $y^{-\beta}$ with $\beta<2$, then $P(k)
\approx k^{\beta/(1-\beta)}$.  As $\beta<2$, the exponent of the
power is smaller than -2.
For example, for $\beta=3/2$ (which
corresponds to the relationship between $x$ and rank $n$ as $x \sim
n^{-2}$ for large $n$, as seen in Fig.1), $P(k) \approx k ^{-3}$ is
obtained.
In general, even if the behavior of $p(y)$ for small $y$ is
not fitted by a power-law, its increase with $y \rightarrow 0$ is slower
than $y^{-2}$.  Then the decrease of $P(k)$ with $k$ is faster than
$k^{-2}$, as often seen in experimental data \cite{barabasi1,barabasi2}.

On the other hand, as $k_{in}$ increases, the
corresponding chemical has greater flow such that the correlation
between the abundance $x$ and $k_{in}$ was formed as in Fig.2c, 
leading to the scale-free network as shown in Fig.2a.

With regard to evolution of reaction networks, preferential attachment
to a more connected node has often been discussed \cite{barabasi1,
barabasi4}.  In contrast, in the present mechanism the attachment is
not preferential at all.  Furthermore, it is different in two
important respects. First, the dynamics of chemical abundance in the
networks were introduced explicitly (described as node `strength' in
\cite{Vespigani}), while previous models generally considered only the
topological structure of the network. Second, selection only by
cellular growth speed results in such a preference, even though
attachment itself is random. Here, we found that more abundant
chemical species acquired more reaction links as attachments of new
links to such chemicals have both a greater influence on the cellular
state and a higher probability of being selected. With these
mechanisms, the power-law in abundance is naturally embedded in the
intracellular reaction network structure through evolution, which is
simply a process of selecting cells with faster growth speeds.

The emergence of two statistical features here is quite general, and
does not rely on the details of our model, such as the kinetic rules
of the reactions or the parameters used. Instead, it is a universal
property of evolution of intracellular reaction dynamics and networks,
selected by growth speed in a fixed environment. The power-laws of
both abundance
and connectivity,
which are often observed in intracellular reactions, are
simply consequences of our mechanism by Darwinian selection.

We would like to thank T. Yomo and K. Sato
for stimulating discussions.
The work is supported by  Grant-in-Aids for Scientific Research from the  
Ministry of Education, Science and Culture of Japan.

\newpage

\noindent
{\Large \bf Figure Captions:}\\
\noindent
{\bf Figure 1:} \\
Rank-ordered
concentration distributions of chemical species. Distributions with
several different generations are superimposed using different
colors. The solid
line indicates the power-law $x \propto n^{-1}$ for the reference. This
power-law of chemical abundance is established around the 10th
generation, and is sustained for further evolutions in the network.
In the simulation, the growth speeds of $10 \times 2000$ networks were
measured, and the top $10$ networks with regards to the growth speed were chosen 
for the next generation.
The parameters were set as $K = 1000$, $D = 4.0$, and $k_{init} = 4$.
~\\

\noindent
{\bf Figure 2:} \\
 Evolution of the network topology.  {\bf (a)}, Connectivity
distribution $P(k)$ of chemical species obtained from the network of the
1000th generation. 
The solid line indicates the power-law $P(k) \propto
k^{-3}$. For comparison, the distribution of $k_{rand}$, obtained by a
randomly generated reaction network with the same number of paths with
the network of 1000th generation, is shown.  {\bf (b)}, Probability $q(x)$ that a
path to a chemical with abundance $x$ is selected in evolution. The
probabilities for incoming ($q_{in}(x)$), outgoing ($q_{out}(x)$), and
catalyzing paths ($q_{cat}(x)$) are plotted. The data were obtained by $1.5
\times 10^5$ trials of randomly adding a reaction path to the network of
the 200th generation, and the paths giving the top 0.05\% growth speeds
were selected.  
~\\

\noindent
{\bf Figure 3:} \\
Changes in the network structure.  The abscissa shows the rank
determined by the abundance of substrate $i$, and the ordinate shows the
rank for the product $j$: the top left is the most abundant and the bottom
right is the least abundant. A point is plotted when there is a reaction
path $i \rightarrow j$, while the abundance of catalyst for the reactions is given
by different colors determined by rank. As each product is dominantly
synthesized from one of the possible paths, we plotted only the path
with the highest flow.  {\bf (a)}, The network at the 10th generation, where
the network structure is rather random, even
though the power-law in abundance has already been established. 
{\bf (b)}, The network at the 1000th generation.  Only a small number of paths
are located in the upper-right triangular portion of the figure,
indicating that almost all chemical species were synthesized from more
abundant species. 

\newpage

\begin{figure}[tbp]
\begin{center}
\includegraphics[width=9.5cm,height=6.5cm]{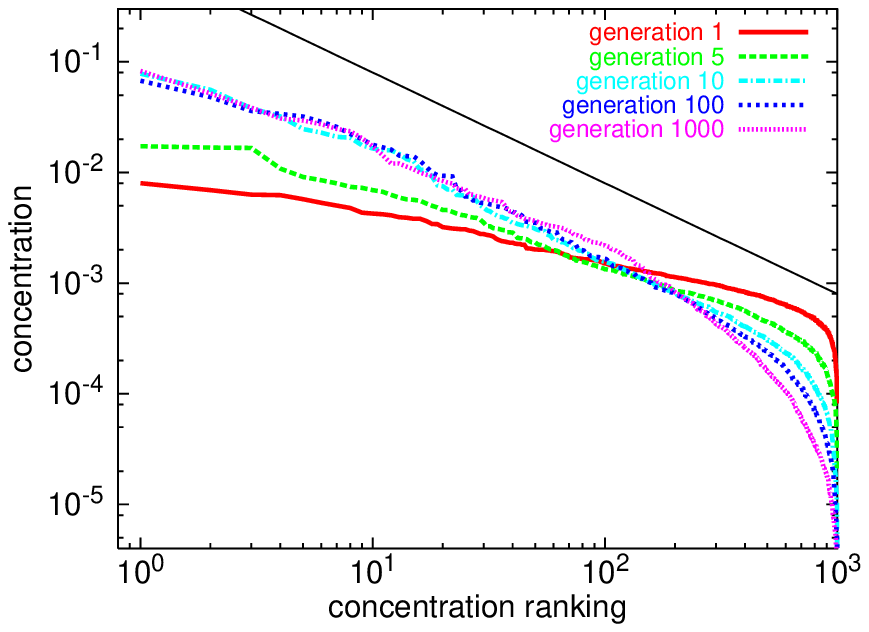}
\caption{
}
\end{center}
\end{figure}

\begin{figure}[htbp]
\begin{center}
\includegraphics[width=9.5cm,height=12.5cm]{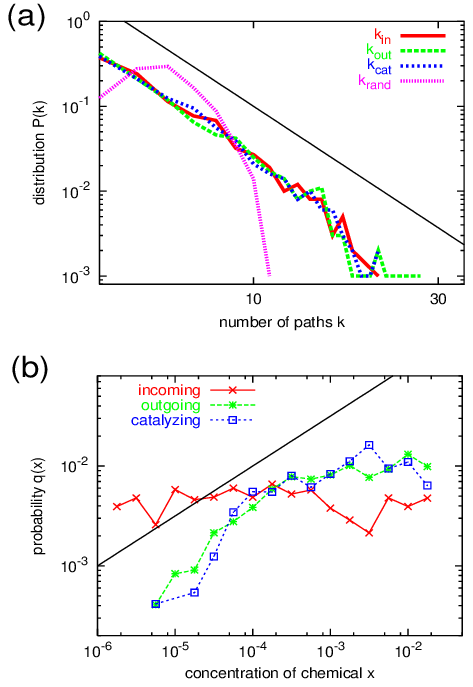}
\caption{}
\end{center}
\end{figure}

\begin{figure}[htbp]
\begin{center}
\includegraphics[width=12cm,height=17cm]{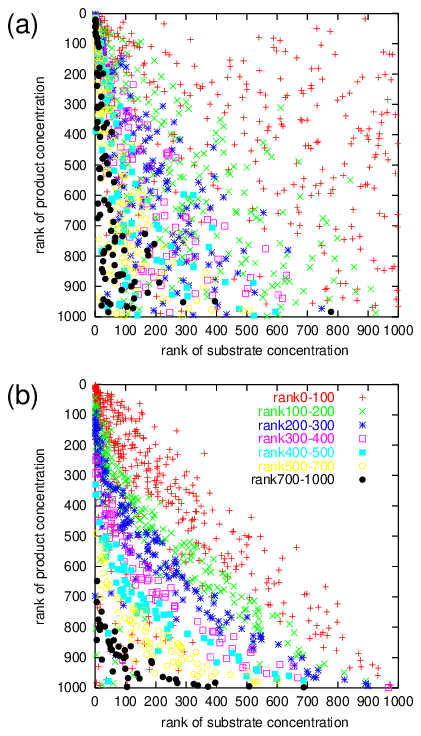}
\end{center}
\end{figure}


\begin{thebibliography}{999}
\bibitem{henkan}
The rank distribution, i.e., the abundances $x$ plotted by rank $n$ 
can be transformed to the density distribution
$p(x)$, 
the probability that the abundance is between $x$ and
$x+dx$. Since \begin{math} dx=dx/dn\times dn \end{math},
there are $|dx/dn|^{-1}$ chemical species between $x$ and $x+dx$.
Thus, if the abundance-rank relation is given by a
power-law with exponent -1, 
$p(x)=|dx/dn|^{-1} \propto n^2 \propto x^{-2}$.
\end{thebibliography}
\end{document}